\documentclass[letterpaper, twoside, 12pt]{article}
\usepackage{amsmath, amssymb, graphicx}
\usepackage[breaklinks, colorlinks, allcolors=blue]{hyperref}

\usepackage[margin=1in]{geometry}

\usepackage[bf]{caption}

\usepackage{fancyhdr}
\usepackage{natbib, ragged2e}
\setcitestyle{aysep={}}

\usepackage{etoolbox}
\makeatletter
\patchcmd{\NAT@citex}
  {\@citea\NAT@hyper@{%
     \NAT@nmfmt{\NAT@nm}%
     \hyper@natlinkbreak{\NAT@aysep\NAT@spacechar}{\@citeb\@extra@b@citeb}%
     \NAT@date}}
  {\@citea\NAT@nmfmt{\NAT@nm}%
   \NAT@aysep\NAT@spacechar\NAT@hyper@{\NAT@date}}{}{}
\patchcmd{\NAT@citex}
  {\@citea\NAT@hyper@{%
     \NAT@nmfmt{\NAT@nm}%
     \hyper@natlinkbreak{\NAT@spacechar\NAT@@open\if*#1*\else#1\NAT@spacechar\fi}%
       {\@citeb\@extra@b@citeb}%
     \NAT@date}}
  {\@citea\NAT@nmfmt{\NAT@nm}%
   \NAT@spacechar\NAT@@open\if*#1*\else#1\NAT@spacechar\fi\NAT@hyper@{\NAT@date}}
  {}{}
\makeatother

\usepackage{titlesec}
\titleformat{\section}{\normalfont\Large\bfseries}{\thesection. }{0pt}{}
\titleformat{\subsection}{\normalfont\large\bfseries}{\thesubsection. }{0pt}{}
\titleformat{\subsubsection}{\normalfont\normalsize\bfseries}{\thesubsubsection. }{0pt}{}

\makeatletter
\def\maketitle{{\centering {\bf \Large \@title \par} \medskip \bf \@author \par }
  \medskip \noindent \@ifnextchar\par{\@gobble}{}}
\makeatother

\newcommand{\hbindex}[1]{#1} 
\newcommand\institute[1]{\relax}
\newenvironment{acknowledgement}{%
  \par \bigskip \noindent \footnotesize \textbf{Acknowledgments.}\space
}{}%

\makeatletter
\let\theorigtitle=\title
\def\title*{\theorigtitle} 
\makeatother

\let\mytablefontsize=\footnotesize

\newcommand\apx{\ensuremath{\sim}}
\newcommand\citeeg[1]{\citep[e.g.,][]{#1}}
\renewcommand\deg{\ensuremath{^\circ}}
\newcommand\fc{\ensuremath{f_\text{C}}}
\newcommand\gbr{G\"udel-Benz relation}
\newcommand\ha{{\ensuremath{\text{H}\alpha}}}
\newcommand\lb{\ensuremath{L_\text{bol}}}
\newcommand\lnur{\ensuremath{L_{\nu,\text{R}}}}

\newcommand\lx{\ensuremath{L_\text{X}}}
\newcommand\mjup{\ensuremath{\text{M}_\text{J}}}
\newcommand\msun{\ensuremath{\text{M}_\odot}}
\newcommand\prot{\ensuremath{\text{P}_\text{rot}}}
\newcommand\rstar{\ensuremath{\text{R}_*}}
\newcommand\speclum{erg~s$^{-1}$~Hz$^{-1}$}
\newcommand\tb{\ensuremath{\text{T}_\text{B}}}
\newcommand\teff{\ensuremath{\text{T}_\text{eff}}}
\newcommand\ujy{\ensuremath{\mu\text{Jy}}}

\newcommand\lnurlb{\ensuremath{\lnur/\lb}}

\newcommand\lxlb{\ensuremath{\lx/\lb}}

\makeatletter
\newcommand\pkgw@simpfx{http://simbad.u-strasbg.fr/simbad/sim-id?Ident=}
\newcommand\MakeObj[4][\@empty]{
  \expandafter\newcommand\csname pkgwobj@c@#2\endcsname[1]{##1}%
  \expandafter\newcommand\csname pkgwobj@f#2\endcsname{#4}%
  \ifx\@empty#1%
    \expandafter\newcommand\csname pkgwobj@s#2\endcsname{#4}%
  \else%
    \expandafter\newcommand\csname pkgwobj@s#2\endcsname{#1}%
  \fi}%
\newcommand{\obj}[1]{%
  \expandafter\ifx\csname pkgwobj@c@#1\endcsname\relax%
    \textbf{[unknown object!]}%
  \else%
    \csname pkgwobj@c@#1\endcsname{\csname pkgwobj@s#1\endcsname}%
  \fi}
\let\obji=\obj
\makeatother

\MakeObj{2m0004}{2MASS\%20J00043484-4044058}{2MASS~J$00043484{-}4044058$~AB}
\MakeObj{2m0024}{2MASS\%20J00242463-0158201}{2MASS~J$00242463{-}0158201$} 
\MakeObj{2m0027}{2MASS\%20J00275592\%2b2219328}{2MASS~J$00275592{+}2219328$~AB} 
\MakeObj{2m0036}{2MASS\%20J00361617\%2b1821104}{2MASS~J$00361617{+}1821104$}
\MakeObj{2m0136}{2MASS\%20J01365662\%2b0933473}{2MASS~J$01365662{+}0933473$} 
\MakeObj{2m0339}{2MASS\%20J03393521-3525440}{2MASS~J$03393521{-}3525440$} 
\MakeObj{2m0423}{2MASS\%20J04234858-0414035}{2MASS~J$04234858{-}0414035$} 
\MakeObj{2m0523}{2MASS\%20J05233822-1403022}{2MASS~J$05233822{-}1403022$}
\MakeObj{2m0607}{2MASS\%20J06073908\%2b2429574}{2MASS~J$06073908{+}2429574$} 
\MakeObj{2m0720}{2MASS\%20J07200325-0846499}{2MASS~J$07200325{-}0846499$~AB}
\MakeObj{2m0746b}{2MASS\%20J07464256\%2b2000321}{2MASS~J$07464256{+}2000321$~B}
\MakeObj{2m0952}{2MASS\%20J09522188-1924319}{2MASS~J$09522188{-}1924319$~AB}
\MakeObj{2m1043}{2MASS\%20J10430758\%2b2225236}{2MASS~J$10430758{+}2225236$}
\MakeObj{2m1047}{2MASS\%20J10475385\%2b2124234}{2MASS~J$10475385{+}2124234$}
\MakeObj{2m1048}{2MASS\%20J10481463-3956062}{2MASS~J$10481463{-}3956062$} 
\MakeObj{wise1122}{WISEP\%20J112254.73\%2b255021.5}{WISEP~J$112254.73{+}255021.5$}
\MakeObj{2m1237}{2MASS\%20J12373919\%2b6526148}{2MASS~J$12373919{+}6526148$}
\MakeObj{2m1314b}{2MASS\%20J13142039\%2b1320011}{2MASS~J$13142039{+}1320011$~B} 
\MakeObj{2m1315}{2MASS\%20J13153094-2649513}{2MASS~J$13153094{-}2649513$~AB}
\MakeObj{2m1456}{2MASS\%20J14563831-2809473}{2MASS~J$14563831{-}2809473$}
\MakeObj{2m1501}{2MASS\%20J15010818\%2b2250020}{2MASS~J$15010818{+}2250020$} 
\MakeObj{2m1835}{2MASS\%20J18353790\%2b3259545}{2MASS~J$18353790{+}3259545$} 
\MakeObj{2m1906}{2MASS\%20J19064801\%2b4011089}{2MASS~J$19064801{+}4011089$} 

\MakeObj{bri0021}{BRI\%20B0021-0214}{BRI~B$0021{-}0214$}
\MakeObj{denis1048}{DENIS\%20J104814.7-395606}{DENIS~J$104814.7{-}395606$}
\MakeObj{lp349}{LP\%20349-25}{LP~$349$--$25$~AB}
\MakeObj{lp944}{LP\%20944-20}{LP~$944$--$20$}
\MakeObj{lsr1835}{LSR\%20J1835\%2b3259}{LSR~J$1835{+}3259$}
\MakeObj{n33370ab}{NLTT\%2033370}{NLTT~$33370$~AB}
\MakeObj{n33370b}{NLTT\%2033370}{NLTT~$33370$~B}
\MakeObj{sdss0423}{SDSS\%20J04234858-0414035}{SDSS~J$04234858{-}0414035$}
\MakeObj{simp0136}{SIMP\%20J01365662\%2b0933473}{SIMP~J$01365662{+}0933473$}
\MakeObj{tvlm513}{TVLM\%20513-46546}{TVLM~$513\text{--}46546$}
\MakeObj{wise0607}{WISEP\%20J060738.65\%2b242953.4}{WISEP~J$060738.65{+}242953.4$}
\MakeObj{wise0720}{WISE\%20J072003.20-084651.2}{WISE~J$072003.20{-}084651.2$}
\MakeObj{wise1906}{WISEP\%20J190648.47\%2b401106.8}{WISEP~J$190648.47{+}401106.8$}

\MakeObj{blcet}{V*\%20BL\%20Cet}{BL~Cet}
\MakeObj{bluvcet}{Gl\%2065}{BL/UV~Cet}
\MakeObj{luh16ab}{NAME\%20Luhman\%2016}{Luhman~16~AB}
\MakeObj{trappist1}{NAME\%20TRAPPIST-1}{TRAPPIST-1}
\MakeObj{uvcet}{V*\%20UV\%20Cet}{UV~Cet}

\begin{document}
\title*{Radio Emission from Ultra-Cool Dwarfs}
\author{P. K. G. Williams}
\institute{P. K. G. Williams \at Harvard-Smithsonian Center for Astrophysics,
  60 Garden Street MS-20, Cambridge, MA 02138, USA \email{pwilliams@cfa.harvard.edu}}

\pagestyle{fancy}
\thispagestyle{plain}
\maketitle

\abstract{\noindent
  \textit{This is an expanded version of a chapter submitted to the Handbook
  of Exoplanets, eds.\ Hans J.~Deeg and Juan Antonio Belmonte, to be published
  by Springer Verlag.}\par The 2001 discovery of radio emission from ultra-cool
  dwarfs (UCDs), the very low-mass stars and brown dwarfs with spectral types
  of \apx M7 and later, revealed that these objects can generate and dissipate
  powerful magnetic fields. Radio observations provide unparalleled insight
  into UCD magnetism: detections extend to brown dwarfs with temperatures
  $\lesssim$1000~K, where no other observational probes are effective. The
  data reveal that UCDs can generate strong (kG) fields, sometimes with a
  stable dipolar structure; that they can produce and retain nonthermal
  plasmas with electron acceleration extending to MeV energies; and that they
  can drive auroral current systems resulting in significant atmospheric
  energy deposition and powerful, coherent radio bursts. Still to be
  understood are the underlying dynamo processes, the precise means by which
  particles are accelerated around these objects, the observed diversity of
  magnetic phenomenologies, and how all of these factors change as the mass of
  the central object approaches that of Jupiter. The answers to these
  questions are doubly important because UCDs are both potential exoplanet
  hosts, as in the TRAPPIST-1 system, \textit{and} analogues of extrasolar
  giant planets themselves. }

\section{Introduction}
\pagestyle{fancy}

The process that generates the solar magnetic field is called
the \textit{\hbindex{dynamo}}. It is widely believed to depend on
the \textit{tachocline}, the shearing layer between the Sun's radiative inner
core and its convective outer envelope \citeeg{c14}. As stellar masses drop
below \apx0.35~\msun\ (spectral types \apx M3.5 and later), the tachocline
disappears \citep{l58, cb00}, which made it challenging to explain how
mid-M~dwarf stars can in fact generate strong magnetic fields \citep{sl85}.

The surprising magnetic properties of fully-convective M~dwarfs raised the
question of what dynamo action would be like in the coolest, lowest-mass
objects: the \textit{\hbindex{ultra-cool dwarfs}} (UCDs), stars and brown
dwarfs with spectral types M7 and later \citep{krl+99, mdb+99}. (The very
youngest and most massive brown dwarfs have spectral types \apx M7; the very
lowest-mass stars have spectral types \apx L4. Objects with spectral types
between these limits can be of either category.) But it was not until the CCD
revolution that it became possible to study UCDs systematically. The first
results suggested that magnetic activity faded out in the UCDs \citeeg{dss+96,
bm95}. The consensus model was that magnetic field generation became
ineffective in the lowest-mass objects due to the loss of the Sun-like
``shell'' dynamo and the transition to cool outer atmospheres, expected to be
largely neutral and therefore unable to couple the energy of their convective
motions into any fields generated below the surface \citep{mbs+02}. This
picture was muddied, however, by reports of flares from very late M~dwarfs in
the ultraviolet \citep[UV;][]{lwb+95}, \ha\ \citep{rkgl99, lkrf99}, and
X-ray \citep{fgs00}. These results suggested that UCDs could generate and
dissipate magnetic fields at least intermittently.

A breakthrough occurred in 2001 with the detection of an X-ray flare
from \obj{lp944}, a \textit{bona fide} brown dwarf \citep[M9.5;][]{rbmb00},
which was shortly followed by the detection of both bursting and quiescent
\hbindex{radio emission} from the same object by a team of summer students using the
NRAO \hbindex{Very Large Array} \citep{bbb+01}. Radio detections of UCDs were
thought to be impossible: scaling arguments had led to radio flux density
predictions of $\lesssim$0.1~\ujy, not achievable even with present-day
observatories. But \citet{bbb+01} detected \obj{lp944} at a flux
density \apx$10^4$ times brighter than these predictions, demonstrating that
UCD magnetism is --- at least sometimes --- vigorous and of a fundamentally
different nature than observed in higher-mass objects. The detection of
quiescent emission further demonstrated that not only can UCDs generate stable
magnetic fields, but that they can also sustainably source the
highly-energetic, nonthermal electrons needed to produce observable radio
emission.

Radio observations have since proved to be the best available probe of
magnetism in the UCD regime, with a major leap in capabilities coming with VLA
upgrade project \citep{pcbw11}. In the rest of this chapter, we describe the
phenomenology of UCD radio emission, place it in a broader astrophysical
context, and deduce the implications of the data for the magnetic properties
of UCDs. We close by presenting the unique contribution that studies of UCD
magnetism can make to exoplanetary science and probable future directions of
research in the field.

\section{Phenomenology of the Radio Emission}

Radio observations of UCDs have revealed a complex phenomenology that can
broadly be divided into ``bursting'' and ``non-bursting'' components. The
non-bursting components can also be variable and evolve significantly over
long timescales (large compared to the rotation period \prot) so we prefer
this use this terminology rather than refer to such emission as ``quiescent.''
\autoref{t.radioucds} presents the list of all known \hbindex{radio-active UCDs} at the
time of writing.

\subsection{Bright, Polarized Bursts}

UCDs emit bright, \hbindex{circularly polarized} radio bursts at GHz
frequencies that have durations $\tau \apx 1$--$100$~minutes. In the initial
discovery by \citet{bbb+01}, the radio bursts of \obj{lp944} had a brightness
temperature $\tb \apx 10^{10}$~K and a fractional circular polarization
$\fc \apx 30$\%, consistent with synchrotron emission mechanisms \citep{d85}.
(Brightness temperature is a proxy for specific intensity often used by radio
astronomers: $I_\nu \equiv 2 \nu^2 k \tb / c^2$.) Subsequent observations
have, however, revealed cases with brightness temperatures and fractional
polarizations too large to be explained by synchrotron emission. In two early
examples, \citet{bp05} detected two bursts from \obj{denis1048} (M8), one with
flux density $S_\nu \apx 20$~mJy, $\tau \apx 5$~minutes, $\tb \apx 10^{13}$~K,
and $\fc \apx 100$\%. \citet{hbl+07} detected repeated bursts
from \obji{tvlm513} (M9) with $S_\nu \apx 3$~mJy, $\tau \apx 5$~minutes,
$\tb \gtrsim 10^{11}$~K, and $\fc \apx 100$\% with both left- and right-handed
helicities observed.

\begin{table}[tbp]
\centering
\def\blend{$^*$}
\mytablefontsize
\begin{tabular}{lllll}
\multicolumn{1}{c}{Source name} &
\multicolumn{1}{c}{Other name} &
\multicolumn{1}{c}{SpT} &
\multicolumn{1}{c}{Var?} &
\multicolumn{1}{c}{First radio detection} \\
\hline

\obj{2m0952} & & M7\blend & & \citet{mbr12} \\
\obj{2m1314b} & \obj{n33370b} & M7 & Y & \citet{mbi+11} \\
\obj{2m1456} & & M7 & & \citet{bp05} \\
\obj{2m0027} & \obj{lp349} & M8\blend & N & \citet{pbol+07} \\
\obj{2m1501} & \obj{tvlm513} & M8.5 & Y & \citet{b02} \\
\obj{2m1835} & \obj{lsr1835} & M8.5 & Y & \citet{b06b} \\
\obj{2m1048} & DENIS~J$\ldots$ & M9 & Y & \citet{bp05} \\
\obj{2m0024} & \obj{bri0021} & M9.5 & Y & \citet{b02} \\
\obj{2m0339} & \obj{lp944} & M9.5 & Y & \citet{bbb+01} \\
\obj{2m0720} & & M9.5+T5 & Y & \citet{bmt+15} \\
\obj{2m0746b} & & L1.5 & Y & \citet{brpb+09} \\
\obj{2m1906} & WISE~J$\ldots$ & L1 & & \citet{gbb+13} \\
\obj{2m0523} & & L2.5 & & \citet{b06b} \\
\obj{2m0036} & & L3.5 & Y & \citet{b02} \\
\obj{2m1315} & & L3.5+T7 & & \citet{bmzb13} \\
\obj{2m0004} & & L5+L5 & & \citet{lmr+16} \\
\obj{2m0423} & SDSS~J$\ldots$ & L7.5 & Y & \citet{khp+16} \\
\obj{2m1043} & & L8 & Y & \citet{khp+16} \\
\obj{2m0607} & WISE~J$\ldots$ & L9 & & \citet{gwb+16} \\
\obj{2m0136} & SIMP~J$\ldots$ & T2.5 & Y & \citet{khp+16} \\
\obj{wise1122} & & T6 & Y & \citet{rw16} \\
\obj{2m1047} & & T6.5 & Y & \citet{rw12} \\
\obj{2m1237} & & T6.5 & Y & \citet{khp+16}

\end{tabular}
\caption{The twenty-three radio-detected UCDs as of mid-2017. ``SpT''
shows a spectral type from SIMBAD; UCD spectral typing is challenging and
subtle \citeeg{kgc+12}, but to conserve space we omit details and references.
Spectral types with asterisks (\blend) are known to come from the blended
spectra of more than one object. ``Var?'' indicates whether the source has
been confirmed to have radio emission that varies on short ($\lesssim$1~hr)
time scales. This is the case for all well-studied UCDs
except \obj{lp349} \citep{opbh+09}.}
\label{t.radioucds}
\end{table}

\begin{figure}[tbp]
\centering
\includegraphics[width=0.8\linewidth]{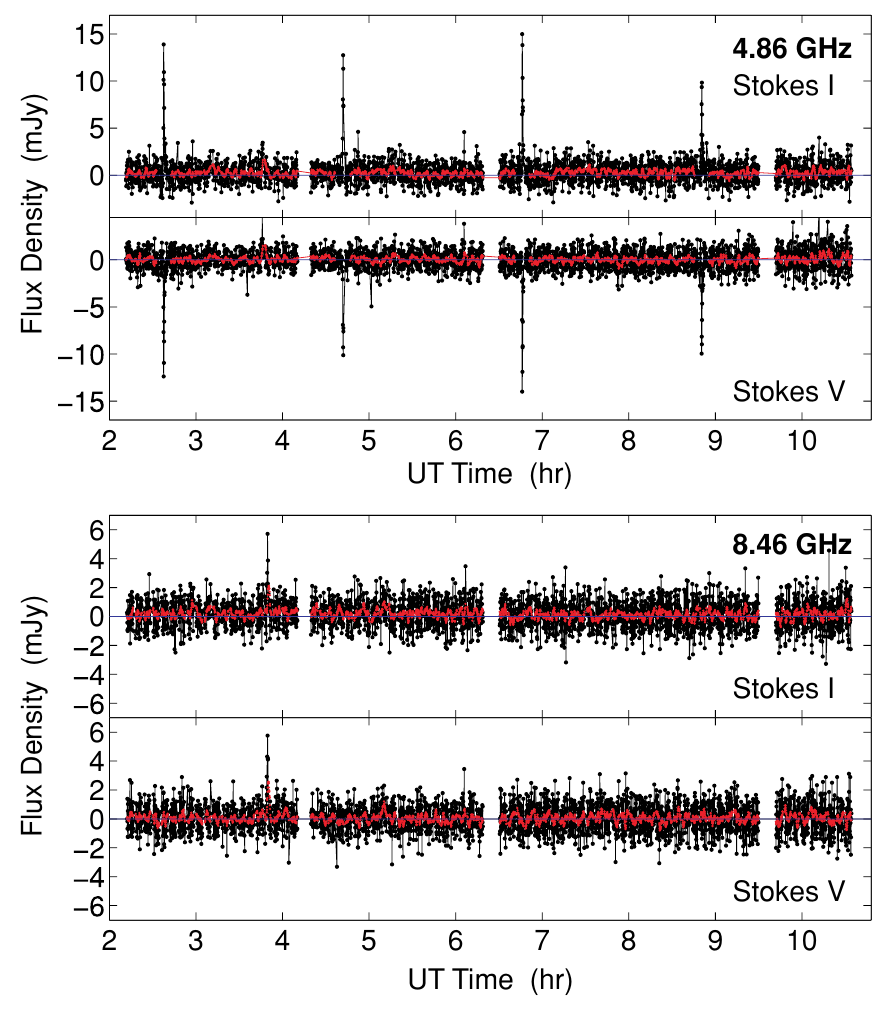}
\caption{\textit{[From \citet{brpb+09}. Reproduced by permission of the AAS.]}
  Radio light curve of \obj{2m0746b} showing periodic, highly polarized, rapid,
  bright bursts. The black and red points show the data averaged into 5- and
  60-second bins, respectively. The negative Stokes~V values, $|V| \apx I$,
  indicate \apx100\% left circular polarization in the bursts. The burst
  spectra do not extend to the VLA's 8.46~GHz band (lower panels).}
\label{f.lightcurve0746}
\end{figure}

In many cases, these radio bursts have been observed to occur periodically,
and in all such cases where the rotation period \prot\ is measured through
independent means, the periodicity of the bursts
matches \prot. \autoref{f.lightcurve0746} shows a classic example of this
phenomenology from \citet{brpb+09}. In the objects with such measurements,
$2 \lesssim \prot \lesssim 4$~h, but there are likely significant selection
effects at play that make it difficult to infer the true distribution
of \prot\ of the radio-active UCDs. In objects with repeated observations, the
periodic bursts are sometimes present and sometimes
not \citep[e.g., \obj{lsr1835};][]{bbg+08, had+08}. \obj{tvlm513} is the
best-studied member of this class, with burst observations spanning years that
enable claims of extremely precise (millisecond) determinations of the
rotation period \citep{dam+10, hhb+13, wr14}.

These bursts have been generally been detected at frequencies between 1 and
10~GHz. Once again, selection effects make it difficult to draw conclusions
about the fundamental character of the burst spectra given the observational
results: the vast majority of searches for UCD radio emission of have been
conducted in the 1--10~GHz frequency window. This window is where the VLA's
sensitivity peaks, but it is challenging to quantify how important intrinsic
effects are as well (we observe in this window because there truly are more
bursts to be seen in it). The spectral shapes of the bursts are not fully
understood. Both high- and low-frequency cutoffs have been observed in
different bursts \citep{lmg15, wbi+15}, but in no burst has there been
definitive evidence that the flux density peak has been identified. Later in
this chapter we will argue that the bursts are probably of moderate bandwidth,
$\Delta\nu / \nu \apx 1$.

The total energy contained in the bursts is not large, which is commonly the
case for radio processes. Using the properties of the bursts
from \obj{tvlm513} quoted above \citep{hbl+07}, the energy content of an
individual burst is \apx$10^{27}$~erg, assuming isotropic emission. For
\hbindex{coherent emission processes} the emission is unlikely to be isotropic, reducing the
energy budget further. The burst luminosities are typically
$\approx${}$10^{-6}$ of the bolometric (sub)stellar radiative output.

\subsection{Non-bursting Emission}

\begin{figure}[tbp]
\centering
\includegraphics[width=\linewidth]{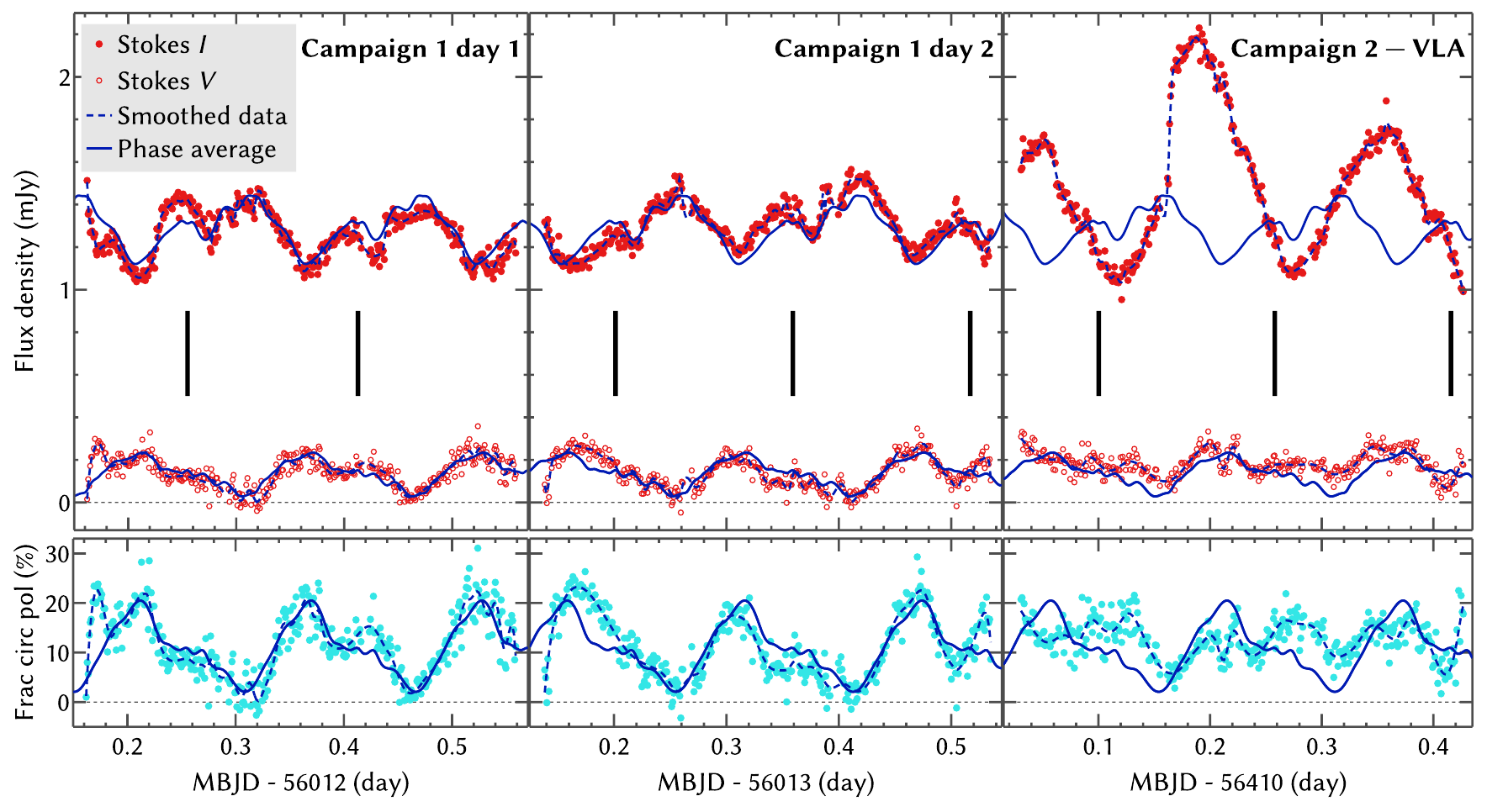}
\caption{\textit{[From \citet{wbi+15}. Reproduced by permission of the AAS.]}
  Radio light curve of \obj{n33370b} showing periodic variation and moderate
  polarization in the non-bursting radio emission. In the upper panels, filled
  and empty points show Stokes~I and V components, respectively. The lower
  panels show the fractional circular polarization derived from these values.
  The leftmost and center panel show two observations separated by 24~hr; the
  rightmost panel shows observations made \apx1~yr later. Vertical black lines
  indicate times that the dwarf's periodically-modulated optical emission
  reaches maximum. Rapid, 100\% circular polarized radio bursts have been
  excised from these data.}
\label{f.lightcurve33370b}
\end{figure}

UCDs also produce non-bursting radio emission that is generally steady over
the time scales of individual observations. Repeated observations of numerous
UCDs have revealed, however, that this emission often varies at the
order-of-magnitude level on longer (\apx week and above)
timescales \citeeg{adh+07, mbr12}. Several UCDs have been detected once in the
radio and not detected in deeper follow-up observations \citeeg{mbr12}. On the
other hand, archival detections show that the hyperactive M7
star \obji{n33370b} has sustained a broadly consistent level of radio emission
for at least a decade \citep{mbi+11}. \autoref{f.lightcurve33370b} shows that
this object, the most radio-bright UCD, nonetheless displays both periodic
(at \prot) and long-term variability in its radio emission.

Radio-detected UCDs typically have non-bursting spectral luminosities of
$\lnur \apx 10^{12}$--$10^{14}$~\speclum, usually about an order of magnitude
fainter than the peak observed burst luminosity when both phenomena have been
observed. Selection effects are important here, too: the lower bound of this
range corresponds to the sensitivity that is achieved in typical VLA
reconaissance observations (\apx1~hr duration) of nearby (\apx10~pc) UCDs. The
deepest upper limit on a UCD is \apx$10^{11}$~\speclum, obtained in
observations of the nearby binary \obji{luh16ab} \citep{oms+15}. The brightest
UCD radio emitter, \obj{n33370b},
reaches \apx$10^{14.7}$~\speclum\ \citep{mbi+11, wcb14}.

The non-bursting emissions generally have low or moderate circular
polarization. Linear polarization has not been detected. As shown
in \autoref{f.lightcurve33370b}, $0 < \fc < 20$\% in the case
of \obj{n33370b}, with periodic variability at \prot\ indicating that the
apparent circular polarization depends on orientation. The recently discovered
radio-active T6.5 dwarf \obj{wise1122} presents a new, unusual case: unlike
the other UCDs, \obj{wise1122} produces highly polarized emission that is not
clearly confined to rapid bursts \citep{wgb17}. The only published
observations of this object are too brief, however, to allow a firm
interpretation.

The non-bursting spectra are broadband. They peak around 1--10~GHz and
generally have shallow spectral indices on both the low- and high-frequency
sides of the peak. Only a few UCDs have been observed at a wide range of radio
frequencies, however. \obj{tvlm513} has been detected at frequencies ranging
from 1.4~GHz all the way to 98~GHz; the latter detection was achieved with
\hbindex{ALMA} and represents the first demonstration that UCDs can be detected at
millimeter wavelengths \citep{wcs+15}. \obj{n33370b} has been detected from
1--40~GHz \citep{mbi+11, wbi+15} and has an extremely flat spectrum, with
significant circular polarization at all observed frequencies. \obj{denis1048}
has been detected from 5--18~GHz with a negative spectral index $\alpha =
-1.71 \pm 0.09$ \citep[$S_\nu \propto \nu^\alpha$;][]{rhhc11}. Searches for
emission from UCDs at frequencies below 1~GHz have thus far been
unsuccessful \citep{jol+11, bhn+16} although the famous low-mass flare
star \obji{uvcet} (M6) was recently detected at 154~MHz using
the \hbindex{Murchison Widefield Array} \citep{llk+17}.

\begin{figure}[tbp]
\centering
\includegraphics[width=0.9\linewidth]{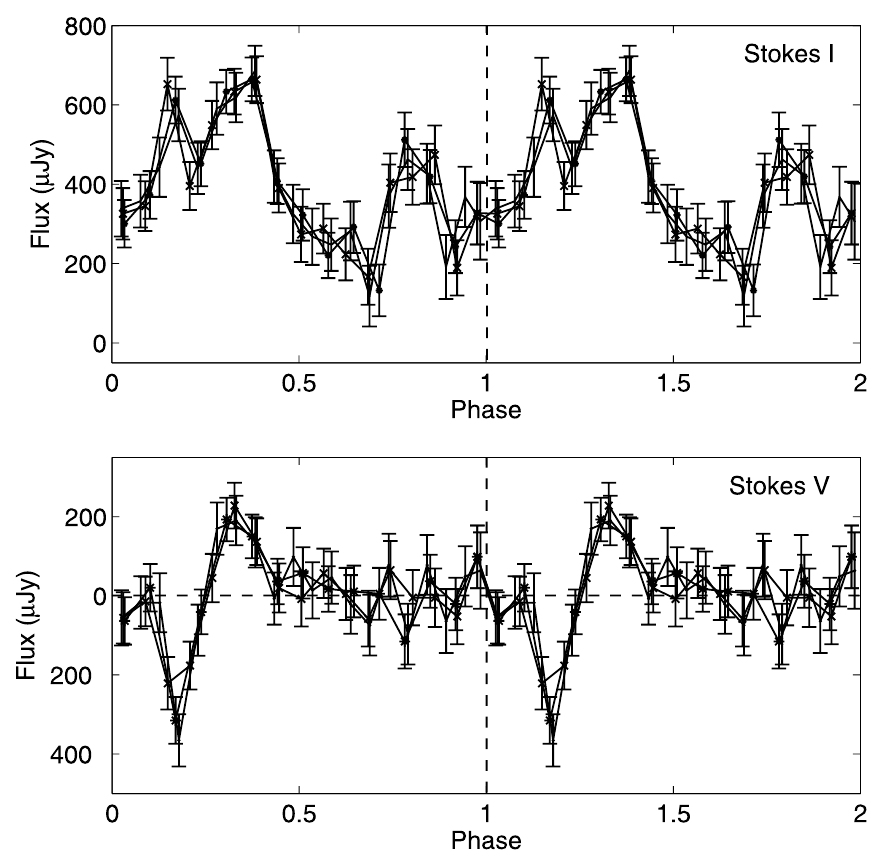}
\caption{\textit{[From \citet{had+06}. Reproduced by permission of the AAS.]}
  Radio light curve of \obj{tvlm513} showing periodic behavior that is not
  cleanly separable into burst and non-burst components. The data are phased
  to a period of 2~hr and shown binned at 6, 7, and 8~minutes, with each
  binned light curve being plotted twice. The observing frequency was
  4.88~GHz.}
\label{f.lightcurve513}
\end{figure}

\subsection{Intermediate Cases}

It is not always possibly to cleanly separate UCD radio emission into bursting
and non-bursting components. \autoref{f.lightcurve513} shows an example
from \obj{tvlm513} in which variability is observed with both circular
polarization helicities as well as null
polarization \citep{had+06}. \obj{2m0036} and \obj{n33370b} have shown
similarly ambiguous phenomenologies \citep{brr+05, had+08, wbi+15}.

\section{UCD Radio Emission in Context}

The previous section focused narrowly on the properties of the radio emission
detected from UCDs. In this section, we place this emission in a broader
astrophysical context.

\subsection{The Prevalence of Radio Activity in UCDs}

Volume-limited \hbindex{radio surveys} of UCDs achieve a detection rate of
approximately 10\% \citep{b06b, mbr12, ahd+13, lmr+16}. However, recent work
by \citet{khp+16} demonstrates that biased surveys can achieve a substantially
higher detection rate: in a sample of five late-L and T~dwarfs selected to
have prior detections of \ha\ emission or \hbindex{optical variability}, four
of the targets were detected. These findings are consistent because the \ha\
detection rate of L and T~dwarfs is also about 10\% on average, with a
noticeably higher detection rate for objects warmer than \apx
L5 \citep{phk+16}.

This ``headline number'' comes with three important caveats. First, it derives
from an observer-dependent binary classification (``did the object's apparent
radio flux density have sufficiently high S/N?'') rather than a fundamental
physical measurement (``what is the object's radio spectral luminosity?'').
Second, the radio detectability of individual objects varies over time in ways
that are not well understood. Third, the reported number averages across a
wide variety of objects, while studies of FGKM dwarfs lead us to expect that
activity strength should depend strongly on fundamental (sub)stellar
parameters.

In particular, mass, rotation, and age are generally believed to be the most
important for setting stellar activity levels \citeeg{b03, wdmh11}.
Correlations between fundamental parameters are pervasive, however, so it is
challenging to determine causation \citeeg{rsp14}. Below we consider how UCD
radio emission scales with some of these physical
parameters, \textit{considering only the radio-detected objects}. A proper
analysis of the entire radio-observed UCD sample that takes into account
nondetections has yet to be performed. Numerous UCDs have upper limits on
their radio emission that are inconsistent with the trends described.

\subsubsection{Mass, Spectral Type, and Effective Temperature}

Because brown dwarfs do not evolve to a stable main sequence and direct mass
measurements of astronomical objects are difficult to obtain, spectral type
(SpT) is widely used as a proxy for mass in UCD activity studies.

The magnetic activity levels of FGKM stars are often quantified with the ratio
of the stellar X-ray luminosity to bolometric luminosity \citep[\lxlb;
e.g.,][]{wdmh11}. This ratio decreases as SpT increases (that is, moves toward
cooler \teff) even though \lb\ on its own scales strongly with \teff, implying
a significant drop in the un-normalized \lx\ \citep{smf+06, bbf+10, wcb14}. It
is therefore striking that in UCDs, \lnur\ shows only a mild decrease with
SpT, with typical values of \apx$10^{13.5}$~\speclum at M7
and \apx$10^{12.5}$~\speclum in the T~dwarfs \citep[their Figure~8]{gwb+16}.
Over this range of SpTs \lnurlb\ increases from typical values
of \apx$10^{-17}$~Hz$^{-1}$ to \apx$10^{-16}$~Hz$^{-1}$.

\subsubsection{Rotation}

Magnetically active FGKM stars follow a ``\hbindex{rotation/activity
relation}'' in which the level of magnetic activity increases with increasing
rotation rate up until a ``\hbindex{saturation} point,'' past which further
increases in rotation rate do not affect the level of magnetic
activity \citeeg{wdmh11}. Here the level of magnetic activity is most commonly
quantified with \lxlb, but analogous trends are observed in most other
measurements that trace activity.

The nature of the radio rotation/activity relationship in UCDs is more
ambiguous. Plots of \lnurlb\ against rotation show a scaling relationship that
has no sign of a saturation point \citep{mbr12}. However, the fastest rotators
tend to be the objects with the latest spectral types, introducing a
covariance with the mass trend described above. \citet{cwb14} studied a subset
of UCDs at a relatively narrow range of SpT, M6.5--M9.5, and found weak
evidence that \lxlb\ is in fact \textit{anti}-correlated with rotation rate.

\subsubsection{Age}

Sun-like stars become less active as they age since they shed angular momentum
through their winds \citep{s72}. This process becomes much less efficient as
stellar mass decreases, with the average activity lifetime of M~dwarfs going
from \apx1~Gyr for M0--M2 stars to \apx8~Gyr for M5--M7 stars \citep{whb+08}.
The data suggest that brown dwarfs rotate rapidly for their entire
lives \citep{bmm+14}.

The relation between age and radio activity has not been studied
systematically in UCDs. However, several noteworthy radio-active UCDs are have
age constraints,
including \obj{lp944} \citep[\apx500~Myr;][]{tr98}, \obj{n33370b} \citep[\apx80~Myr;][]{dfr+16},
and \obji{simp0136} \citep[\apx200~Myr;][]{gfb+17}. Very young UCDs can also
have radio emission associated with young-star phenomena such as accretion,
jets, and disks \citeeg{rzp17}.

\begin{figure}[tbp]
\centering
\includegraphics[width=0.9\linewidth]{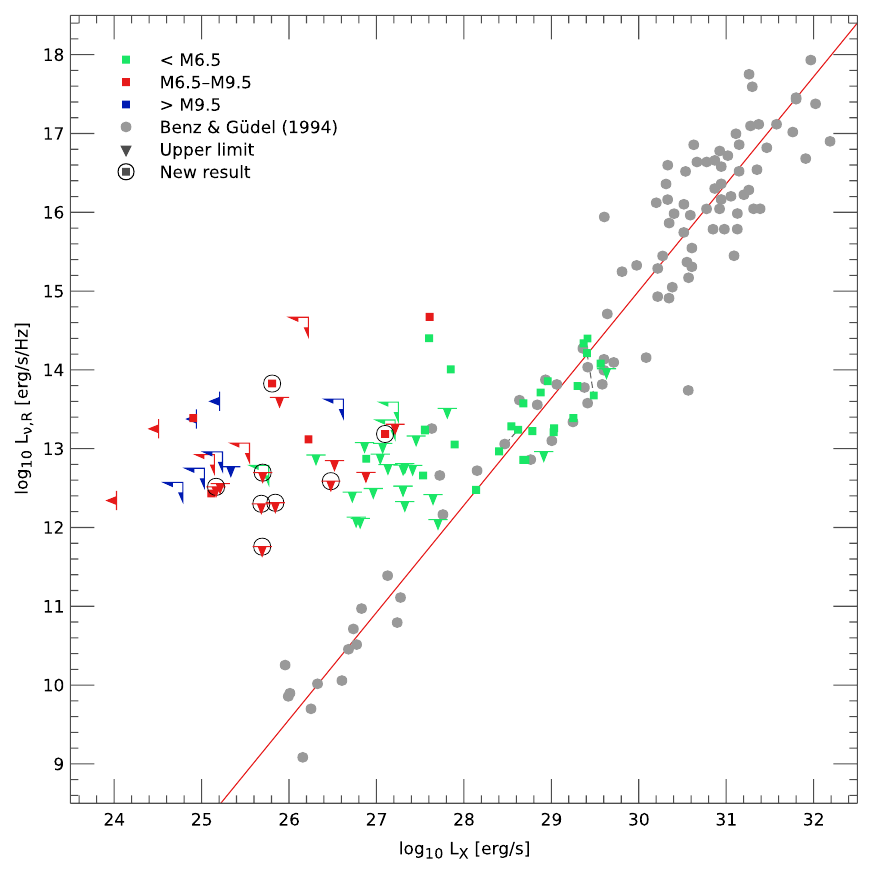}
\caption{\textit{[From \citet{wcb14}. Reproduced by permission of the AAS.]}
  Radio and X-ray emission for active stars and brown dwarfs. Gray points and
  the red line show the ``\gbr'' defined for active stars and solar flares.
  Green, red, and blue points show data for M3--M6, M6.5--M9.5, and $\ge$L0
  dwarfs, respectively. While some UCDs may obey the \gbr, there is a
  substantial population of outliers with radio emission that far exceeds what
  would be predicted from their X-ray emission.}
\label{f.lrlx}
\end{figure}

\subsection{Multi-wavelength Correlations}

Stellar magnetism is associated with emission across the electromagnetic
spectrum, and different bands probe different physical regions or processes.
In Sun-like stars, \ha\ emission probes the chromosphere; UV, the transition
region; X-rays, hot dense coronal plasma; and radio/millimeter emission,
particle acceleration. Multi-wavelength observations, especially simultaneous
ones, therefore yield insights that cannot be obtained through single-band
studies.

The radio and X-ray luminosities of active stars are nearly linearly
correlated, a phenomenon known as the ``\hbindex{G\"udel-Benz
relation}'' \citep{gb93, bg94}. A single power law can fit observations
spanning ten orders of magnitude in \lnur, in systems ranging in size from
individual solar flares to active binaries (\autoref{f.lrlx}, gray points). As
shown above, however, the \gbr\ breaks down dramatically in the UCD
regime \citep[\autoref{f.lrlx}, colored points;][]{bbb+01, wcb14}.
Correlations between the luminosities of UCDs in radio and other bands
(e.g., \ha) have not yet been investigated in the literature.

\hbindex{Simultaneous multi-wavelength observations} can illuminate the physics of
stellar and substellar flares, although extensive observations of flare stars
demonstrate that very few general statements can be made: individual events
may or may not be associated with emission in each of the bands that trace
magnetic activity, and the relative ordering and magnitude of the emission in
these bands is variable \citeeg{oba+04}. The UCD with the best simultaneous
multi-wavelength observational coverage is \obj{n33370b}, and the data show a
similar variety of phenomenologies \citep{wbi+15}. A detailed understanding of
the underlying physics remains elusive.

The evidence that optical/IR variability is a useful indicator of UCD radio
activity \citep{khp+16} suggests that the two are correlated. Only a handful
of UCDs have data sets that allow the optical and radio variability (either
bursts or non-bursting periodic variatons) to be phased. While the radio and
optical maxima of \obj{tvlm513} are significantly out of phase \citep{wr14,
mzpop15}, there is a hint that the millimeter and optical maxima may occur at
the same phase \citep{wcs+15}. The radio and optical maxima of \obj{n33370b}
are also significantly out of phase. Intriguingly, long-term monitoring of
this object suggests that its non-bursting polarized radio emission remains in
phase with its optical variability but the total radio intensity does
not \citep{wbi+15}. \textit{Keck} spectroscopic monitoring of \obji{lsr1835}
revealed periodic variations in the optical emission that were argued to
originate in a high-altitude opaque blackbody with $T \apx
2200$~K \citep{hlc+15}.

Radio and \ha\ variability can also be correlated. In \obj{2m0746b}, the radio
bursts are 90\deg\ out of phase with the maxima of periodic changes in
the \ha\ equivalent width \citep{brpb+09}. Recent observations
of \obj{lsr1835} showed radio and \ha\ variations that were approximately in
phase \citep{hlc+15}, but other observations of the same object have shown
aperiodic \ha\ variability with no clear connection to the radio
emission \citep{bbg+08}. Simultaneous multi-wavelength monitoring
of \obj{tvlm513} revealed periodic \ha\ variability with no clear connection
to emission in other bands, although there is some evidence for radio bursts
at the times of the \ha\ minima \citep{bgg+08}.

\section{Interpretation of the Data}

We now turn to the astrophysical interpretation of the observations presented
in the previous sections.

\subsection{Auroral Radio Emission}

The periodic, bright, highly-polarized radio bursts observed in radio-active
UCDs are consistent with the \hbindex{auroral radio bursts} observed in Solar
System planets \citep{ztrr01}, which are generally agreed to originate from
the
\hbindex{electron cyclotron maser instability} \citep[ECMI;][]{wl79, t06}. The ECMI
converts the free energy of a magnetized plasma into electromagnetic waves
through resonant interactions between the waves and the particles' cyclotron
motion. The ECMI is relatively easy to trigger in physical systems involving
beams of mildly relativistic electrons that are accelerated along magnetic
field lines by the presence of a co-aligned electric field, if the ambient
medium is of sufficiently low density. This happens at the Earth when
energetic solar wind particles funnel down its magnetic field lines toward the
poles.

Observable ECMI emission is expected to be dominated by a narrow-band signal
at the electron cyclotron frequency of the local magnetic field,
\begin{equation}
\nu_\text{ce} = \frac{e B}{2 \pi m_e c}
  \apx 2.8 \left(\frac{B}{1\text{ G}}\right)\text{ MHz}.
\end{equation}
Observations of ECMI bursts from UCDs therefore measure the strengths of their
magnetic fields. In practice, the ECMI occurs in regions that span a variety
of field strengths, so the observed emission has a moderate bandwidth,
$\Delta \nu / \nu \apx 1$ \citep{ztrr01}, with a cutoff at high frequencies
because the body's magnetic field reaches some peak value at its surface. ECMI
emission is beamed and likely refracts through the plasmasphere that evidently
envelops the radio-active UCDs, necessitating detailed simulations to predict
its observed properties \citeeg{kdy+12, ydk+12}.

\begin{sloppypar} At a typical VLA observing frequency of \apx5~GHz, the
inferred strength is \apx2~kG, comparable to the strongest \hbindex{surface
field strengths} observed on active M~dwarfs \citep{kps+17}. A polarized pulse
at 10~GHz from the T6.5 dwarf \obj{2m1047} implies a field strength of at
least 3.6~kG \citep{wb15}, demonstrating that the fully convective dynamo can
generate strong fields even in extremely low-mass, cool (\apx$900$~K) objects.
\end{sloppypar}

Observations of multiple consecutive ECMI bursts at the rotation period imply
the presence of a relatively stable ``\hbindex{electrodynamic engine}'' that
accelerates the beams of electrons responsible for the emission. Understanding
the nature of this engine is one of the great tasks in the field of UCD
magnetism. In the Solar System planets, the engine is often powered by the
solar wind \citeeg{d61, a69}, but this driver is not available for solivagant
UCDs. The only persuasive explanation is that the engine is ultimately powered
by the body's rotation \citep{s09}. This is largely the case for
\hbindex{Jupiter} \citep{mb07}, raising the exciting possibility that sophisticated
models developed in the context of the Solar System gas giants can be brought
to bear on the UCD case. For instance, studies of Jupiter inform a model in
which rotational energy is converted into nonthermal particle acceleration
through shear-induced currents at the corotation breakdown
radius \citep{nbc+12}.

Rapid rotation and the stable operation of the electrodynamic engine imply
that the magnetospheres of radio-active UCDs likely have dipole-dominated
topologies. This inference is supported by observations that probe the
topologies of the magnetic fields of cool stars. Studies using \hbindex{Zeeman
Doppler Imaging}
\citep[ZDI;][]{the.zdi} show that strong, axisymmetric, dipolar
fields emerge in the coolest M~dwarfs currently accessible to the
technique \citep{mdp+10} and that such a topology may be associated with
enhanced radio activity and variability \citep{kl17}.

Auroral electron beams do not only produce radio emission. First, auroral
processes are associated with emission across the electromagnetic spectrum,
with the highest luminosities concentrated at FUV and IR
wavelengths \citep{bg00}. However, the emission at these wavelengths is not
nearly as bright as it is in the radio, such that the auroral fluxes in other
bands inferred for known active UCDs are beyond the capabilities of
present-day instruments. Second, the energetic auroral electrons eventually
precipitate into the upper atmosphere, where they can drive chemical processes
like \hbindex{haze production} \citeeg{wyf03}. \citet{hlc+15} interpreted
their simultaneous radio and optical observations in this framework, arguing
that an electron beam delivering $10^{24}\text{--}10^{26}$~erg~s$^{-1}$ of
kinetic power drove both the radio emission of \obj{lsr1835} and its optical
variability by creating a compact, high-altitude layer of H$^{-}$ upon
precipitation. This model also motivated the targeted survey
of \citet{khp+16}, under the assumption that auroral electron beams cause
detectable \ha\ and/or optical variability. A recent study, however, does not
find a correlation between \ha\ and high-amplitude optical variability in a
large sample of L/T dwarfs \citep{mmha17}.

\subsection{Gyrosynchrotron Radio Emission}

Non-bursting UCD radio emission bears the hallmarks
of \hbindex{gyrosynchrotron emission}, the same process that is believed to be
responsible for the bulk of the radio emission observed from active
stars \citep{d85, g02}. Gyrosynchrotron emission is produced by mildly
relativistic electrons spiraling in an ambient magnetic field, resulting in a
broadband spectrum with low to moderate circular polarization. Analysis of the
spectral properties can constrain the ambient magnetic field strength, the
total number and volume density of energetic particles, and their energy
distribution. It has been argued that the non-bursting UCD radio emission may
instead represent an unusual form of ECMI emission \citep{had+06, had+08}, but
several lines of evidence, most notably the millimeter-wavelength detection
of \obj{tvlm513}, discourage this interpretation \citep{wbi+15, wcs+15}.

The standard equations for gyrosynchrotron emission are derived for spatially
homogeneous field and particle properties \citep{d85}. A robust result of this
analysis is that the optically-thick (low-frequency) side of the spectrum
should have a spectral index $\alpha = 5/2$, much steeper than that observed
for sources like \obj{n33370b} and \obj{tvlm513} \citep{ohbr06, mbi+11}. While
the flat observed spectra can be reproduced qualitatively with more realistic
inhomogeneous models \citeeg{wkj89, tlu+04}, homogeneous models should still
give a sense of the average properties of the emitting region. Spectral fits
with both kinds of model suggest that the ambient field strength in the
synchrotron-emitting region is \apx$10$--$100$~G, typical of flare
stars \citep{b06b, ohbr06, mka+17}. Assuming standard energetic electron
densities and brightness temperatures, the typical source size is a
few \rstar\ \citep{b06b, wcb14}. The fact that \rstar\ evolves only slowly
with mass in the UCD regime may help explain why \lnur\ appears to settle at a
typical value of \apx$10^{13}$~\speclum\ in the radio-active UCDs, if the
other factors that set the synchrotron radio luminosity ($B$ and $n_e$) are
also mass-insensitive. Radio emission is energetically insignificant, so if
the particle acceleration process saturates in some way, this value of \lnur\
could be achieved in UCDs with widely varying bolometric and spindown
luminosities.

Analyses of the non-bursting radio emission of UCDs have not yet begun to
leverage the detailed models that have been developed for analogous systems.
\hbindex{Magnetic chemically peculiar (MCP) stars} have high masses but also possess
strong, dipole-dominated magnetospheres with persistent and periodically
variable radio emission. Numerical modeling of MCP particle populations can
constrain the magnetospheric structure in detail \citep{tlu+04, lto+17}. Even
more excitingly, Jupiter's \hbindex{radiation (van~Allen) belts} have been
studied in exquisite detail and produce centimeter-wavelength emission with
variability, spectra, and polarization that are highly reminiscent of the UCD
observations \citep{dp81, dpbg+03}. The application of Jovian models to UCD
data has the potential to yield a treasure trove of insight. For instance, the
presence of Jupiter's moons can be inferred from the spectrum of its radiation
belts alone \citep{scb08}, and observations made at different orientations of
the planet can be combined to reconstruct the full three-dimensional structure
of the belts \citep{sodl97}.

\subsection{The Emergence of ``Planet-like'' Magnetism in UCDs}

The data show that UCDs can generate strong magnetic fields and dissipate
their energy vigorously, but that they do so in processes that are
fundamentally different than the typical flare star phenomenology. This is
demonstrated most clearly by the substantial drop in UCD X-ray emission
(both \lx\ and \lxlb), violation of the \gbr, and the emergence of periodic,
bright, highly-polarized radio bursts.

This can be understood as the emergence of ``planet-like'' magnetism in UCDs,
characterized by processes that occur in large-scale, stable,
rotation-dominated magnetospheres \citep{s09}. These include the operation of
an electrodynamic engine that accelerates auroral electron beams and sustains
a population of mildly relativistic electrons. The lack of X-ray emission
indicates that coronal heating, if it can be said to occur at all, does not
happen in a Sun-like fashion. Historically, this has been explained as being
due to the outer atmosphere becoming electrically neutral and therefore unable
to couple the energy of convective motions into magnetic flux
tubes \citep{mbs+02}. More recent work has argued that UCD atmospheres should
in fact still couple to the magnetic field efficiently \citep{rhsr15},
suggesting that more detailed analysis is needed.

One of the fundamental questions about this picture is why only \apx10\% of
UCDs are detected in the radio. While early thinking focused on the possible
roles of inclination and rotation rate \citeeg{hhk+13}, current data suggest
that planet-like magnetism is only \textit{sometimes} present in UCDs and that
the presence or absence of planet-like behavior is not linked to any
particular fundamental parameter. The most compelling evidence for this is
the \obj{n33370ab} system: while \obj{n33370b} is the most radio-luminous UCD
known, its binary companion is at least 30 times fainter than it, despite
being nearly identical in mass, age, rotation rate, and
composition \citep{wbi+15, dfr+16, fdr+16}. Population studies show evidence
for bimodality when considering the \gbr\ \citep{sab+12, wcb14}, the
rotation/activity relation \citep{cwb14}, and ZDI-derived magnetic field
topologies \citep{mdp+10}.

The large-scale topology of the magnetic field may be the key factor that
determines whether planet-like magnetic behavior arises in a given
UCD \citep{cwb14}. This hypothesis is tenable because \hbindex{geodynamo
simulations} indicate that the fully convective dynamo may be bistable in the
conditions encountered in the UCD regime, with identical objects sustaining
different topologies depending on initial conditions \citep{gmd+13}. Recent
observations provide the first direct evidence for this model: ZDI reveals
that \obj{uvcet} (\apx M6) has an axisymmetric, dipole-dominated magnetic
field, while the field of its nearly-identical binary companion \obj{blcet} is
weaker and non-axisymmetric \citep{kl17}. Consistent with the proposed
model, \obj{uvcet} is more luminous and variable in the radio
than \obj{blcet}.

Detectable radio emission requires the presence of both a magnetic
field \textit{and} nonthermal electrons. The difference between the
radio-active and -inactive UCDs may therefore hinge not on the field topology
but on the presence of a source of plasma that can eventually produce the
gyrosynchrotron and ECMI emission. In analogy with Jupiter, the 10\% of UCDs
that are radio-active might be the ones possessing volcanic planets resembling
\hbindex{Io}. This scenario can potentially be tested by searching for ECMI bursts that
repeat periodically not at \prot\ but at the synodic period of the planetary
orbit. No evidence of such a non-rotational periodicity has yet been reported.

\section{The Exoplanetary Connection}

Radio studies of UCDs make a unique contribution to exoplanetary science
because they are the only effective way to observe the magnetic properties of
cool, extrasolar bodies.

\begin{sloppypar}
One reason that this is important is that UCDs may host large numbers of
observationally-accessible small planets, as demonstrated by
the \obji{trappist1} system \citep{gjl+16, gtd+17}. Understanding UCD activity
is therefore important for the same reasons that it is important for any
exoplanet host star: magnetic phenomena make planet discovery more
challenging \citeeg{rmer14} and they can have a significant impact on
\hbindex{atmospheric retention} and the broader question of
\hbindex{habitability} \citeeg{jglb15, sbj16}. Because UCD magnetism can be
so different from that of Sun-like stars and M~dwarfs, its impact on
habitability may differ substantially from the cases that have been
investigated thus far in the literature. For instance, the detection of
millimeter-wavelength radiation from \obj{tvlm513} points to a surprisingly
high-energy radiation environment of MeV electrons, which can
produce \hbindex{$\gamma$-ray emission} when they precipitate into the stellar
atmosphere \citep{wcs+15}. The moons of the Solar System gas giants should
serve as useful reference points in this domain \citeeg{ppw08}.
\end{sloppypar}

UCD magnetic fields can strongly resemble those of the Solar System gas giant
planets. Radio observations therefore provide insight into the magnetospheres
of exoplanets themselves, which observers have been struggling to probe since
well before the first confirmed exoplanet discovery \citep{yse77}. Currently,
\hbindex{exoplanetary magnetospheres} can only be investigated using indirect and
model-dependent means \citeeg{ehw+10}. Direct observations of exoplanetary
magnetospheres would not only shed light on the question of habitability, but
also internal structure; for instance, magnetic field generation in rocky
planets may require the presence of plate tectonics \citep{bls10}. The first
direct measurement of the magnetic field of a planetary-mass object may
already have occurred, because \obj{simp0136}, detected in the radio
by \citet{khp+16}, was recently argued to be a member of the \apx200-Myr-old
Carina-Near moving group, which would give it a mass of $12.7 \pm 1.0$~\mjup\
according to standard evolutionary models \citep{gfb+17}.

\section{Future Directions of Research}

One of the top priorities in the field of UCD radio studies is the extension
of its techniques to genuine exoplanets. By analogy with Solar System
examples, exoplanets are expected to have magnetic fields that are much weaker
than those of UCDs, which leads to the expectation that their radio emission
will occur at lower radio frequencies, $\lesssim$300~MHz. Fortunately the past
decade has witnessed a dramatic investment in low-frequency radio arrays such
as the \hbindex{Low Frequency Array} (LOFAR), the \hbindex{Murchison Widefield Array} (MWA), the
\hbindex{Long-Wavelength Array} (LWA), the \hbindex{Giant Metrewave Radio Telescope} (GMRT), and
the \hbindex{Hydrogen Epoch of Reionization Array} (HERA). While the first
generation of these instruments has not yielded any detections of genuine
UCDs, the first positive results are starting to emerge \citep{llk+17}, and
virtually all of these observatories are undergoing upgrades that are expected
to yield significant sensitivity improvements.

While many of the nearest UCDs have been surveyed by the Very Large Array, the
results of \citet{khp+16} suggest that targeted searches may be able to yield
detections beyond the typical detection horizon (\apx30~pc) for blind searches
thus far. Furthermore, radio studies of southern UCDs have historically been
hampered by the lack of an instrument as powerful as the VLA (latitude
$+34$\deg). The commissioning of the \hbindex{MeerKAT} radio telescope in South
Africa \citep{the.meerkat}, with science operations slated to begin in late
2017, will introduce a powerful new observatory in the south. MeerKAT should
be especially valuable in surveys for radio emission from young,
\hbindex{directly-imaged exoplanets}, which are promising targets because they are as
warm, or even warmer, than the coolest UCDs with confirmed radio detections,
and convect vigorously. Most of the currently-known young planets are in the
southern hemisphere, however, and have not been the subject of sensitive radio
observations. Surveys for radio-active UCDs in both hemispheres will be
transformed by the deeper insight into the natures of the stars and brown
dwarfs in the solar neighborhood afforded by upcoming surveys from
observatories such as Gaia, the Transiting Exoplanet Survey Satellite (TESS),
and Spektr-RG, the spacecraft bearing the \hbindex{e-ROSITA} instrument.

Finally, a great deal of theoretical work remains to be done. More detailed
models of the bursting and non-bursting radio emission will strengthen the
astrophysical inferences that can be drawn from the radio data. The
populations statistics of radio-active UCDs should be understood better by a
more rigorous treatment of the many nondetections and a more careful
characterization of the long-term variability of their radio emission. This
sort of work will lay the foundations upon which models can be constructed
that explain fundamental puzzles such as the source of the radio-emitting
plasma, the possible existence of a bistable dynamo, and the relationship
between rotation and magnetic activity in the ultra-cool regime.

\begin{acknowledgement}
P.K.G.W. thanks Edo Berger for supporting his work in this field and Adam
Burgasser, Kelle Cruz, Trent Dupuy, Jackie Faherty, Gregg Hallinan, Mark
Marley, and Rachel Osten for many enlightening conversations over the years.
P.K.G.W. acknowledges support for this work from the National Science
Foundation through Grant AST-1614770. This research has made use of the SIMBAD
database, operated at CDS, Strasbourg, France and NASA's Astrophysics Data
System.
\end{acknowledgement}

\bibliography{\jobname}

\end{document}